\documentclass[11pt]{article}
\usepackage{amsmath}
\usepackage{amssymb}
\usepackage{amsmath,amsthm,epsfig,euscript,array,cite,cancel}
\usepackage{mathtools}

\newcommand{\be}{\begin{equation}}
\newcommand{\bea}{\begin{eqnarray}}
\newcommand{\ee}{\end{equation}}
\newcommand{\eea}{\end{eqnarray}}

\begin{document}
\topmargin -1cm \oddsidemargin=0.25cm\evensidemargin=0.25cm
\setcounter{page}0
\renewcommand{\thefootnote}{\fnsymbol{footnote}}
\begin{titlepage}
\begin{flushright}
MPP-2014-2
\end{flushright}
\vskip .7in
\begin{center}
{\LARGE \bf Higher Spins in Hyperspace \\

 }  \vskip .6in {\Large Ioannis Florakis$^a$\footnote{e-mail: {\tt florakis@mppmu.mpg.de}},
Dmitri Sorokin$^b$\footnote{e-mail: {\tt  dmitri.sorokin@pd.infn.it }} and
 Mirian Tsulaia$^c$\footnote{e-mail: {\tt  mirian.tsulaia@canberra.edu.au}  }}
\vskip .4in {$^a$ \it Max-Planck-Institut f\"ur Physik, Werner-Heisenberg-Institut, F\"ohringer Ring 6, 80805 M\"unchen, Germany} \\
\vskip .2in {$^b$ \it INFN, Sezione di Padova, via F. Marzolo 8, 35131 Padova, Italia} \\
\vskip .2in { $^c$ \it Faculty of  Education Science Technology and Mathematics, University of
Canberra, University Dr, Bruce ACT 2617, Australia}\\
\vskip .8in

\begin{abstract}

We consider the $Sp(2n)$ invariant formulation of higher spin fields on flat and curved backgrounds of constant curvature.
In this formulation an infinite number of higher spin fields are packed into single scalar and spinor master fields (hyperfields) propagating on extended spaces, to be called hyperspaces, parametrized by tensorial coordinates.
We show that the free field equations on flat and AdS--like hyperspaces are related to each other by a generalized conformal transformation
of the scalar and spinor master fields. We compute the four--point functions on a flat hyperspace
for both scalar and spinor master fields, thus extending the two-- and three--point function results of  hep-th/0312244. Then using
the generalized conformal transformation we derive two--, three-- and four--point functions on AdS--like hyperspace from the corresponding correlators on the flat hyperspace.

\end{abstract}

\end{center}

\vfill

\end{titlepage}

\renewcommand{\thefootnote}{\arabic{footnote}}
\setcounter{footnote}0

\section{Introduction} \setcounter{equation}0
Various formulations of one and the same theory may prove useful for revealing and/or making manifest its different properties and features. This is certainly the case for higher spin gauge theory, for which various different descriptions have been proposed. Historically, the first approach was a metric--like formulation put forward by Fronsdal \cite{Fronsdal:1978rb} and the second one was the frame--like approach \cite{Vasiliev:1980as,Aragone:1980rk} which proved to be most efficient for constructing non--linear
higher spin field
theories with the use of unfolding techniques \cite{Vasiliev:1990en,Vasiliev:1992av,Vasiliev:2003ev} \footnote{See reviews \cite{Vasiliev:1999ba,Sorokin:2004ie,Bekaert:2005vh,Bouatta:2004kk,Fotopoulos:2008ka,Bekaert:2010hw,Sagnotti:2011qp,Tsulaia:2012rb} and references therein for details on the features and different formulations of higher spin theory.}.  Since consistent interactions require an infinite number of fields with spin ranging from zero to infinity, in this approach,
massless higher spin gauge fields are encoded into a generalized (one--form) spin connection and a scalar (zero--form) field which, in the case of four space--time dimensions $x^m$ $(m=0,1,2,3)$, take the following form
\bea \label{int-omega}
\omega(x,y,\overline y)= \sum_{i,j=0}^\infty dx^m
\omega_m^{\alpha_1 ... \alpha_i, \,\dot \beta_1 \dots \dot \beta_j} (x) y_{\alpha_1}...y_{\alpha_i}
 {\overline y}_{\dot \beta_1}...
{\overline y}_{\dot \beta_j},\nonumber\\
C(x,y,\bar y)=\sum_{i,j=0}^\infty C^{\alpha_1 ... \alpha_i, \,\dot \beta_1 \dots \dot \beta_j} (x) y_{\alpha_1}...y_{\alpha_i}
 {\overline y}_{\dot \beta_1}...
{\overline y}_{\dot \beta_j}\,,
\eea
where $y_\alpha$ and $\bar y_{\dot \beta}$ $(\alpha,\dot\beta=1,2)$ are twistor--like Weyl--spinor variables, which are used to
\if{}
which obey commutation relations.
\begin{equation}
[y_\alpha, y_\beta]= \epsilon_{\alpha \beta}, \quad [y_{\dot \alpha}, y_{\dot \beta}]= \epsilon_{{\dot \alpha} {\dot \beta}}.
\end{equation}
\fi
incorporate into a compact form an infinite number of physical higher spin gauge fields and their field strengths with spins $s$
growing from zero to infinity, as well as an infinite number of auxiliary fields. The variables $y_\alpha$ and $\bar y_{\dot \beta}$ can be regarded as coordinates that extend the conventional space--time with additional `twistor--like' directions.
\if{}
In (\ref{int-omega}) the physical fields are described by the
  $\omega_m^{\dot \alpha_1, ... \dot \alpha_{s-1}, \dot \beta_1, ,,,\dot \beta_{s-1}} (x)$
components of the generalized spin connection whereas the other components are auxiliary fields.
\fi
The interacting theory is formulated in space--time with a non--zero cosmological constant $\Lambda$, for instance on a $D$--dimensional anti de Sitter background. The coupling constants in a perturbative expansion of higher spin interactions are proportional
to inverse powers of $\Lambda$ and thus do not admit a naive flat space--time limit\footnote{Although Vasiliev's nonlinear equations
are formulated in a background independent way, their perturbative expansion is also (usually) performed around the $AdS_D$ background which is one of the admissible vacuum solutions of these equations, see \emph{e.g.} \cite{Vasiliev:1999ba,Bekaert:2005vh,Bekaert:2010hw} for a review.}.
The theory is gauge invariant under an infinite--dimensional non--Abelian higher spin gauge symmetry, which contains
an $AdS_D$ isometry group $SO(D-1,2)$ as a finite dimensional subgroup. In $D=4$ and in the free field limit, the $SO(3,2)$ symmetry is extended to the conformal symmetry $SO(4,2)$, the latter being (spontaneously) broken by higher spin interactions.
\if{}
In the metric like approach the higher spin fields are described by tensor fields of an arbitrary rank
$\phi_{\mu_1,..., \mu_s}(x)$. The free field equations and Lagrangians for the higher spin fields possess a certain amount of gauge invariance,
which is enough to remove nonphysical polarizations, the ghosts, from the system.
As a second step one makes a nonlinear deformation of these Abelian gauge transformations and also deforms the field equations
with nonlinear terms in such a way that the gauge invariance is maintained order by order in the coupling constant.
This procedure can be carried out on both Minkowski and Anti-de Sitter spaces, although one should further carry out extra consistency checks for a flat background, whereas the computations on AdS background are technically more complicated than in the frame
like- approach.
\fi
Nevertheless, it is often important to first understand the symmetries of the free theory, which are already quite nontrivial for higher spin gauge theories, and then study their implications when the interactions are switched on.
The requirement that the nonlinear interaction possesses a part of, or some kind of nonlinear deformations of the original free theory, can be a good selection criterion for the allowed interaction terms.
The study of one of the ``hidden" symmetries of free field equations of massless higher spin fields and, in particular,
the restrictions that this symmetry imposes on their correlation functions in flat and  AdS spaces
is the subject of this paper. In $D=4$, the hidden symmetry in question is $Sp(8)$ and contains the conformal group $SO(4,2)$ as a subgroup. To make the $Sp(8)$ symmetry manifest, we will consider a formulation of free higher spin theory in which the conventional space--time is extended with extra coordinates in a way that is different from (or complementary to) eq. \eqref{int-omega}. We will call such an extension `\emph{hyperspace}' \cite{Vasiliev:2007yc} to reconcile the different names given in earlier papers, such as \emph{tensorial space} \cite{Bandos:1998vz,Bandos:1999qf} or \emph{matrix space} \cite{Vasiliev:2001dc}.

The symmetry that we are going to explore was first observed in \cite{Fronsdal:1985pd} using  the following reasoning.
It is well known that the group $SO(3,2)\sim Sp(4)$, which is  the isometry group  of a four--dimensional AdS space
and of the conformal group in three dimensions, has a so-called singleton representation associated with a $3d$ scalar and spinor field.
According to the Flato--Fronsdal theorem \cite{Flato:1978qz},
the $Sp(4)\times Sp(4)$ product of two singleton modules generates an infinite sum of massless higher spin states in $D=4$ with each spin $s$ appearing once.  The integer and half--integer spin sets of these states form infinite representations of the  $Sp(8)$ group, which contains
$Sp(4)\times Sp(4)$ and $SO(4,2)$ as subgroups. Fronsdal observed that ten is the minimal dimension of space, which contains the four--dimensional space--time as a subspace, and in which the $Sp(8)$ symmetry acts geometrically, i.e. it acts on the points of this space in a way similar to conformal transformations in flat or AdS space--time. His idea was that there should exist a theory in this $10d$ hyperspace which, in a way alternative to that of Kaluza and Klein, would reproduce the massless higher spin field theory in the $4d$ space--time.

The first explicit realization of this idea was a twistor--like superparticle model of Bandos and Lukierski
\cite{Bandos:1998vz} which, for $D=4$, possesses the generalized superconformal symmetry under $OSp(1|8)$. The original motivation behind this model
was not related to higher spins, but to a geometric interpretation of commuting tensorial charges of an extended supersymmetry algebra
as momenta conjugate to six tensorial coordinates $y^{mn}=-y^{nm}$ $(m,n=0,1,2,3)$, which extend four space--time coordinates $x^m$ to the ten--dimensional hyperspace
\be\label{X}
X^{\mu\nu}=X^{\nu\mu}=\frac 12\,x^m\,\gamma_m^{\mu\nu}+\frac 14
\,y^{mn}\,\gamma_{mn}^{\mu\nu}\ , \qquad
\mu,\nu=1,2,3,4\,,
\ee
where $\gamma_m^{\mu\nu}=\gamma_m^{\nu\mu}$ are four--dimensional symmetric gamma--matrices.

The higher spin content of this model was found
later in \cite{Bandos:1999qf} where the quantum
states of the superparticle were shown to form an infinite tower of
massless higher spin fields, and the relation of this model to the unfolded formulation was assumed. This relation was analyzed in
detail in \cite{Vasiliev:2001zy,Vasiliev:2001dc,Plyushchay:2003gv,Plyushchay:2003tj,Bandos:2005mb}.

In particular, in \cite{Vasiliev:2001zy} it was demonstrated that
 the field equations in a super--hyperspace $\mathcal M_{N|n}$ of  bosonic dimension $\frac 12 n(n+1)$ and of
fermionic dimension $nN$ are $OSp(N|2n)$ invariant and, for $n=4$, they
correspond to the unfolded higher spin field free equations in
$D=4$. It has also been shown \cite{Vasiliev:2001dc} that the theory
possesses properties of causality and locality.
A detailed analysis of free field equations in hyperspaces associated with  space--times of dimension $D=3,4,6$ and $10$
was further carried out in \cite{Bandos:2005mb}.
Two-- and three--point $Sp(2n)$--invariant correlation functions of scalar and spinor fields in flat hyperspace $\mathcal M_{n}$ were computed in \cite{Vasiliev:2001dc,Vasiliev:2003jc}. In the unfolded formalism, $Sp(2n)$--invariant $multi$--point functions were given in \cite{Didenko:2012tv}, that generalized four--dimensional 2-- and 3--point function computations of \cite{Colombo:2012jx}.
Other aspects of the hyperspace formulation and its supersymmetrization have been considered in  \cite{Gelfond:2003vh,Bandos:2004nn,Vasiliev:2007yc,Ivanov:2007vx,Gelfond:2008ur,Gelfond:2008td,Gelfond:2010pm,Bandos:2011wi}
(see also \cite{Fedoruk:2012ka}).

The results mentioned above were obtained in flat hyperspace that contains conventional Minkowski space--time as a subspace. However, also $AdS$ (super)spaces admit the hyperspace extensions \cite{Bandos:1999pq,Bandos:1999rp}, \cite{Vasiliev:2001zy}. These are (super)group manifolds $OSp(N|n)$. In particular, the hyperspace extension of $N=1$ $AdS_4$ superspace is the supergroup $OSp(1|4)$. In \cite{Didenko:2003aa} and \cite{Plyushchay:2003tj} it was shown that $Sp(8)$--invariant field equations on $Sp(4)$ lead to free unfolded equations for massless higher spin fields in $AdS_4$.

In this paper, we continue the study of the dynamics of massless higher spin fields in flat and $Sp(n)$ hyperspaces. In particular, exploiting the property that $Sp(n)$ group manifolds are `GL--flat' \cite{Plyushchay:2003gv,Plyushchay:2003tj}, \emph{i.e.} they are related to the flat hyperspace by a `generalized conformal' (general linear) transformation, we find the explicit relation between the solutions of the $Sp(2n)$--invariant field equations in flat hyperspace and on $Sp(n)$, as well as the relation between the $Sp(2n)$--invariant correlation functions of fields in these spaces. Requiring $Sp(2n)$ symmetry, we also derive the explicit form of the four--point correlation functions in these hyperspaces, which turns out to be analogous to the form of correlation functions in conformal field theories.

The paper is organized as follows. In  Section \ref{Hyperflat}, we collect the main facts about scalar and spinor field  theories on flat hyperspaces. We give an explicit form of   the field equations, describe their  $Sp(2n)$ symmetry group and review how  the linearized curvatures for massless higher spin fields in the conventional flat space--time are obtained in this approach.

In Section  \ref{sechypads}, we discuss the scalar and spinor field theories on  $Sp(n)$ group manifolds. As mentioned above, these manifolds are actually hyperspace extensions of AdS spaces, and the field equations on $Sp(n)$ manifolds are deformations of the ones on flat hyperspaces, with the deformation
parameter being related to the corresponding AdS radius.

In  Section \ref{secgl4}, we establish a connection between the previous two Sections. In particular, we show that the field
equations on flat and AdS hyperspaces are related via a generalized conformal transformation of the scalar and spinor fields, similarly
to the case of scalar and spinor fields on the ordinary flat and AdS spaces.
The crucial tool in establishing the connection between flat and AdS hyperspaces is
the $GL(n)$  flatness property of $Sp(n)$ group manifolds \cite{Plyushchay:2003gv},
which is a generalization of the conformal flatness property
of conventional AdS spaces.

Section \ref{secads} is, in a certain sense, complementary to the rest of the paper. There,
we show by explicit computation how the metric on a four dimensional AdS space is obtained
from the $Sp(4)$ hyperspace and derive the exact relation between
the contraction parameter of the $Sp(4)$ algebra and the radius of $AdS_4$.

In Section \ref{seccf} we present computations of various correlation functions on flat and AdS hyperspaces.
 The two-- and three-- point correlation functions  on flat hyperspaces were obtained previously in \cite{Vasiliev:2003jc}.
We follow a similar approach to derive four--point functions on flat hyperspace  for bosonic and fermionic fields.
Having obtained correlators on flat hyperspaces, we use the generalized conformal transformation relating the  fields on flat
and AdS hyperspaces in order to obtain the correlators on $Sp(n)$ group manifolds.

The last Section contains our conclusions and open questions for future research.

Finally, the Appendix summarizes technical details that are useful for the calculations.

\section{Scalar and spinor field theory in flat hyperspace} \setcounter{equation}0 \label{Hyperflat}
The points of the flat hyperspace ${\mathcal M}_{n}$ are parametrized by symmetric matrix coordinates $X^{\mu\nu}=X^{\nu\mu}$ $(\mu,\nu=1,\ldots,n)$. The linear symmetries of ${\mathcal M}_{n}$ are rigid translations and $GL(n)$ rotations generated, respectively, by
\be\label{Pmunu}
P_{\mu\nu}=-i\frac{\partial}{\partial X^{\mu\nu}}\equiv-i\partial_{\mu\nu}\,,\qquad [P_{\mu\nu},\,P_{\rho\lambda}]=0\,,
\ee
and
\be\label{GL}
L_{\nu}{}^{\mu}=-2i X^{\mu\rho}\,\partial_{\rho\nu}\,, \qquad [L_{\nu}{}^{\mu},L_{\lambda}{}^{\rho}]= i(\delta^{\mu}_\lambda\,L^{\rho}{}_{\nu}-\delta^{\rho}_\nu\,L^{\mu}{}_{\lambda})\,,
\ee
where, by definition,
\begin{equation}\label{derivative}
\frac{ \partial X^{\mu \nu}}{\partial X^{\rho \lambda}} = \frac{1}{2}(\delta^\mu_\rho \delta^\nu_\lambda +
\delta^\nu_\rho \delta^\mu_\lambda)\,.
\end{equation}
Under \eqref{Pmunu} and \eqref{GL} the hyperspace coordinates are transformed as follows
\be\label{deltaXP}
\delta X^{\mu\nu}=i(a^{\rho\lambda }P_{\rho\lambda}+g_{\rho}{}^{\lambda}\,L_{\lambda}{}^{\rho}) X^{\mu\nu}= a^{\mu\nu} + (X^{\mu\rho}g_\rho{}^\nu+X^{\nu\rho}g_\rho{}^\mu)\,,
\ee
where $a^{\mu\nu}=a^{\nu\mu}$ and $g_{\mu}{}^\nu$ are arbitrary constant parameters.

These symmetries are the hyperspace counterparts of the conventional Poincar\'e translations, Lorentz rotations and dilatations of  Minkowski space--time. Generalized Lorentz rotations are generated by traceless operators $L_{\mu}{}^{\nu}-\frac 1n \delta^{\nu}_\mu\,L_{\lambda}{}^\lambda$, forming the $SL(n)$--algebra, whereas dilatations are generated by the trace of $L_{\mu}{}^{\nu}$.

One may enlarge these transformations by considering generalized conformal boosts
\be\label{K}
K^{\mu\nu}=iX^{\mu\rho}X^{\nu\lambda}\partial_{\rho\lambda}\,,\qquad [K^{\mu\nu},\,K^{\rho\lambda}]=0\,,
\ee
so that the total transformation of $X^{\mu\nu}$ becomes
\be\label{deltaX}
\delta X^{\mu\nu}=i(a^{\rho\lambda }P_{\rho\lambda}+g_{\rho}{}^{\lambda}\,L_{\lambda}{}^{\rho} + k_{\rho \lambda}K^{\rho \lambda}) X^{\mu\nu}= a^{\mu\nu} + (X^{\mu\rho}g_\rho{}^\nu+X^{\nu\rho}g_\rho{}^\mu)-X^{\mu\rho}k_{\rho\lambda}X^{\lambda\nu}\,,
\ee
where $k_{\mu\nu}=k_{\nu\mu}$ are constant parameters of the boosts.

The generators \eqref{Pmunu}, \eqref{GL} and \eqref{K} form the $Sp(2n)$ algebra which plays the role of a generalized conformal symmetry in the hyperspace
\bea\label{sp2n}
&[P_{\mu\nu},\,P_{\rho\lambda}]=0,\qquad [K^{\mu\nu},\,K^{\rho\lambda}]=0,\qquad [L_{\nu}{}^{\mu},L_{\lambda}{}^{\rho}]= i(\delta^{\mu}_\lambda\,L_{\nu}{}^{\rho}-\delta^{\rho}_\nu\,L_{\lambda}{}^{\mu})\,,&\nonumber \\
&[P_{\mu \nu},L_\lambda{}^\rho]=-i(\delta_\mu^\rho P_{\nu \lambda} + \delta^\rho_\nu P_{\mu \lambda} ),\qquad
 [K^{\mu \nu},L_\lambda{}^\rho]=i  ( \delta^\mu_\lambda  K^{\nu \rho}   + \delta^\nu_\lambda  K^{\mu \rho}  )\,,&\nonumber \\
& [P_{\mu \nu},K^{\lambda \rho}]=\frac{i}{4} (  \delta^\rho_\mu L_\nu{}^\lambda +  \delta^\rho_\nu L_\mu{}^\lambda
+  \delta^\lambda_\mu L_\nu{}^\rho +  \delta^\lambda_\nu L_\mu{}^\rho)  \,. &
\eea
From the structure of this algebra, one can see that the flat hyperspace $\mathcal M_n$ can be realized as a coset manifold associated with the translations $P=\frac {Sp(2n)}{K\times\!\!\!\!\supset SL(n)}$  where $K\times\!\!\!\!\!\!\supset SL(n)$ is the semi--direct product of the general linear group and the boosts $K_{\mu\nu}$.

In the case $n=4$, which is related to the higher spin theory in $D=4$ (see eq. \eqref{X}), the generalized conformal symmetry of $\mathcal M_{4}$ is $Sp(8)$.
As was previously shown in \cite{Vasiliev:2001zy}, the dynamics of the free higher spin fields in flat $D=4$ space--time is encoded into two hyperfields. A scalar field $b(X)$ incorporates the field strengths of the $4d$ fields of integer spins and a spinor field $f_\mu(X)$
incorporates the half--integer spin field strengths\footnote{We call the field $f_\mu(X)$ spinor since in the physically interesting cases the index $\mu$ is associated with a spinor representation of the Lorenz group in the $D$--dimensional subspace--time of the hyperspace. }. They satisfy the following field equations \cite{Vasiliev:2001zy}
\begin{equation}\label{BF}
(\partial_{\mu \nu} \partial_{ \rho \lambda}- \partial_{\mu \rho} \partial_{\nu \lambda}) b(X)=0\,,
\end{equation}
\begin{equation}\label{FF}
\partial_{\mu \nu}f_\rho(X) -\partial_{\mu \rho}f_\nu(X)=0\,.
\end{equation}
Note that, in the above equations, there is no contraction of indices, implying that \emph{a priori} we do not endow the hyperspace with a metric structure. As we will see below, the metric structure will appear upon reduction of these equations to the physical space--time by expanding the tensorial coordinates in the basis of the gamma--matrices as in eq. \eqref{X}. The Minkowski metric then appears as a consequence of the use of the Clifford algebra $\{\gamma^{m},\gamma^n\}=2\eta^{mn}$.

In any $\mathcal M_n$, the equations \eqref{BF} and \eqref{FF} are invariant under the $Sp(2n)$ transformations \eqref{deltaX}, provided that the fields transform as follows
\begin{equation}\label{sp8fb}
\delta b(X)= -(a^{\mu \nu} \partial_{\mu \nu} + \frac{1}{2}g_\mu{}^\mu + 2 g_\nu{}^\mu X^{\nu \rho}
\partial_{\mu \rho} -
 k_{\mu \nu} (\frac{1}{2} X^{\mu \nu}   + X^{\mu \rho} X^{\nu \lambda}\partial_{\rho \lambda})) b(X)\,,
\end{equation}

\begin{eqnarray}\label{sp8ff} \nonumber
\delta f_\rho(X)&=& -(a^{\mu \nu} \partial_{\mu \nu} + \frac{1}{2}g_\mu{}^\mu + 2 g_\nu{}^\mu X^{\nu \lambda}
\partial_{\mu \lambda} -
 k_{\mu \nu} (\frac{1}{2} X^{\mu \nu}   + X^{\mu \tau} X^{\nu \lambda}\partial_{\tau \lambda})) f_\rho(X) +
\\
&&-(g_\rho{}^\nu - k_{\lambda \rho}X^{\lambda \nu}) f_\nu(X)\,.
\end{eqnarray}
Note that these variations contain the term $\frac{1}{2}(g_\mu{}^\mu-k_{\mu \nu} X^{\mu \nu})$,  implying that the fields have the canonical conformal weight $1/2$. A natural generalization of these transformations for fields of a generic conformal weight $\Delta$ is
\begin{equation}\label{sp8fbD}
\delta b(X)= -(a^{\mu \nu} \partial_{\mu \nu} + \Delta\,(g_\mu{}^\mu-
 k_{\mu \nu}  X^{\mu \nu}) + 2 g_\nu{}^\mu X^{\nu \rho}
\partial_{\mu \rho} -
 k_{\mu \nu}  X^{\mu \rho} X^{\nu \lambda}\partial_{\rho \lambda}) b(X)\,,
\end{equation}

\begin{eqnarray}\label{sp8ffD} \nonumber
\delta f_\rho(X)&=& -(a^{\mu \nu} \partial_{\mu \nu} +\Delta\,(g_\mu{}^\mu-
 k_{\mu \nu}  X^{\mu \nu}) + 2 g_\nu{}^\mu X^{\nu \lambda}
\partial_{\mu \lambda} -
 k_{\mu \nu} X^{\mu \tau} X^{\nu \lambda}\partial_{\tau \lambda}) f_\rho(X)
\\
&&-(g_\rho{}^\nu - k_{\lambda \rho}X^{\lambda \nu}) f_\nu(X)\,.
\end{eqnarray}

In the case of $n=2$, the hyperspace ${\mathcal M}_2$ is just the ordinary $D=3$ Minkowski space parametrized by $X^{\mu\nu}=x^m\,\gamma_m^{\mu\nu}$ $(m=0,1,2)$ and, as one may easily check, eqs. \eqref{BF} and \eqref{FF} reduce, respectively, to the Klein--Gordon equation for the massless scalar $b(x)$ and the massless Dirac equation for the Majorana spinor $f_\mu(x)$, which are conformally invariant.

In the case of $\mathcal M_4$, eqs. \eqref{BF} and \eqref{FF} produce in $D=4$ the conformally invariant set of Bianchi identities and equations of motion for linearized field strengths of the massless fields of all spins $s=0,\frac 12 , 1, 2, \dots, \infty$, while the cases $n=8$ and $n=16$ describe conformally invariant higher spin fields whose field strengths are self--dual, respectively, in $D=6$ and $D=10$, as was shown in detail in \cite{Bandos:2005mb}.

For instance, to obtain the higher spin field equations from \eqref{BF} and \eqref{FF} in the four--dimensional case,  one expands $b(X)$ and $f_\mu(X)$ in powers of the extra coordinates $y^{mn}=-y^{nm}$, eq. \eqref{X}, as follows
\begin{eqnarray}\label{ymn}
b(x^l,\,y^{mn})&=\phi(x)+y^{m_1n_1}F_{m_1n_1}(x)
+y^{m_1n_1}\,y^{m_2n_2}\,[R_{m_1n_1,m_2n_2}(x)-{1\over 2}\eta_{m_1m_2}\partial_{n_1}\partial_{n_2}\phi(x)]\nonumber\\
&+\sum_{s=3}^{\infty}\,y^{m_1n_1}\cdots
y^{m_sn_s}\,[R_{m_1n_1,\cdots,m_sn_s}(x)+\cdots]\,,\nonumber\\
~&\\
 f^\rho(x^l,y^{mn})
 &{\hspace{-150pt}}
=\psi^\rho(x)+y^{m_1n_1}[{ R}^\rho_{m_1n_1}(x)-{1\over
2}\partial_{m_1}(\gamma_{n_1}\psi)^\rho]\nonumber\\
& +\sum_{s={5\over 2}}^{\infty}\,y^{m_1n_1}\cdots y^{m_{s-{1\over
2}}n_{s-{1\over 2}}}\,[{R}^\rho_{m_1n_1,\cdots,m_{s-{1\over
2}}n_{s-{1\over 2}}}(x)+\cdots]\,.\nonumber
\end{eqnarray}
In (\ref{ymn}), $\phi(x)$ and $\psi^\rho(x)$ are a scalar and a spinor
field, respectively, $F_{m_1n_1}(x)$ is the Maxwell field strength,
$R_{m_1n_1,m_2n_2}(x)$ is the curvature tensor of linearized
gravity, ${ R}^\rho_{m_1n_1}(x)$ is the Rarita--Schwinger field
strength and other terms in the series stand for generalized Riemann
curvatures of spin--$s$ fields\footnote{The pairs of the indices separated by the commas are antisymmetrized.}
(that also contain contributions of
derivatives of the fields of lower spin denoted by dots, as in the case of
the Rarita--Schwinger and gravity fields).

Substituting the expressions \eqref{ymn} into  eqs. \eqref{BF} and \eqref{FF}, and rewriting the derivatives explicitly as
\be
\partial_{\mu\nu}=\frac 12 \left( \gamma^m_{\mu\nu}\frac\partial{\partial x^m}+\gamma^{mn}_{\mu\nu} \frac\partial{\partial y^{mn}}
\right)\,,
\ee
one finds that the scalar and the spinor field satisfy,
respectively, the Klein--Gordon and the Dirac equations, while the
higher spin field curvatures satisfy the Bianchi identities
\begin{equation}\label{B1}
R_{[m_1n_1,\,m_2]n_2,\cdots,\,m_sn_s}=0\,,\qquad \partial_{[l_1}R_{m_1n_1],\,m_2n_2,\cdots,\,m_sn_s}=0\,,
\end{equation}
and the linearized higher spin field
equations
\begin{equation}\label{trbis}
R^m{}_{n_1,mn_2,m_3n_3,\cdots,\,m_sn_s}=0\,, \qquad (\gamma^{m_1}{R})^\mu_{m_1n_1,\,m_2n_2,\,\cdots\,,m_{s-{1\over
2}}n_{s-{1\over 2}} }=0\,.
\end{equation}
To the best of our knowledge, equations similar to \eqref{B1} and \eqref{trbis} first appeared in the Weinberg paper \cite{Weinberg:1965rz}. In \cite{deWit:1979pe}, in a `symmetric' Young--tableaux convention, higher spin curvatures (and generalized Christoffel symbols) were constructed as $s$--derivatives of Fronsdal \cite{Fronsdal:1978rb} potentials of spin--$s$.  In the same form as \eqref{B1} and \eqref{trbis} the curvature equations were given \emph{e.g.} in \cite{Howe:1988ft}, and in  \cite{Bekaert:2002dt,Bekaert:2003az} it was shown that these equations for integer--spin curvatures are equivalent to s--derivative equations on unconstrained spin--s potentials and are invariant under unconstrained local higher--spin symmetries. These equations, in turn, are reduced (upon a partial gauge fixing) to the second--order Fronsdal equations \cite{Bekaert:2003az}. For half--integer higher--spin fields these results were generalized in \cite{Bandos:2005mb} to which we refer the reader for further details on field theories in flat hyperspaces and proceed to discuss hyperspace field theories related to higher spin fields in $AdS$.

\section{Scalar and spinor field theory on the group manifold $Sp(n)$} \label{sechypads}\setcounter{equation}0
As was noticed in  \cite{Bandos:1999pq,Bandos:1999rp} and \cite{Vasiliev:2001zy}, the hyperspace extension of the $AdS_4$ space is the group manifold $SO(3,2)\sim Sp(4)$ which contains the $AdS_4=\frac{SO(3,2)}{SO(3,1)}$ symmetric space as a coset subspace of maximal dimension. For $n>4$, an $AdS_d$ space is also a subspace of $Sp(n)$ but is no longer the maximal coset of this group.

Before generalizing the field equations \eqref{BF} and \eqref{FF} to the $Sp(n)$ case, let us recall the  basic group--theoretical and geometric properties of the $Sp(n)$ group manifold.

The group $Sp(n)$ is generated by $n \times n$ symmetric matrices $M_{\alpha \beta}$ forming the algebra
\begin{equation}\label{algebrasp8}
\left[ M_{\alpha \beta}, M_{\gamma \delta} \right] = - \frac{i\xi}{2} \left[ C_{\gamma (\alpha} M_{\beta) \delta}+
 C_{\delta (\alpha} M_{\beta) \gamma} \right], \quad \alpha, \beta= 1,...,n\,,
\end{equation}
where $C_{\alpha\beta}=-C_{\beta\alpha}$ is an $Sp(n)$--invariant symplectic metric and the parameter $\xi$  has the inverse dimension of length. As will be shown explicitly in Section \ref{secads}, the parameter $\xi$ is related to the  radius of
the AdS space. Its presence in the $Sp(n)$ algebra allows one to perform (at $\xi\rightarrow 0$) its contraction  to the algebra of translations $M_{\alpha \beta}\rightarrow P_{\alpha\beta}$ \eqref{Pmunu} of the flat ${\mathcal M}_n$ hyperspace.

As a group manifold, $Sp(n)$ is the coset  $Sp(n)_L\times Sp(n)_R/Sp(n)$ which has the isometry group $Sp(n)_L\times Sp(n)_R$, the latter being the subgroup of $Sp(2n)$ generated by
\be\label{LR}
M^L_{\alpha\beta}= P_{\alpha\beta}-\frac{\xi^2}{16}K_{\alpha\beta}- \frac{\xi}{4}L_{(\alpha\beta)}\,\qquad M^R_{\alpha\beta}= P_{\alpha\beta} -  \frac{\xi^2}{16}  K_{\alpha\beta} +  \frac{\xi}{4} L_{(\alpha\beta)}\,,
\ee
as one may see from the structure of the $Sp(2n)$ algebra \eqref{sp2n}. In \eqref{LR}, $K_{\alpha\beta}=C_{\alpha\gamma}C_{\beta\delta}K^{\gamma\delta}$ and $L_{(\alpha\beta)}=\frac 12(L_{\alpha}{}^{\gamma}C_{\gamma\beta}+L_{\beta}{}^{\gamma}C_{\gamma\alpha})$. The latter generate the diagonal $Sp(n)$ subalgebra of $Sp(n)_L\times Sp(n)_R$. This algebraic structure implies that $Sp(n)$ can also be realized as a coset manifold of $Sp(2n)$ associated with the generators $P-\xi^2 K=\frac {Sp(2n)}{SL(n)\times\!\!\!\!\supset K}$. This coset is apparently different from the $Sp(2n)$ coset realization of the flat
hyperspace ${\mathcal M}_n$ discussed in the previous Section, but it implies that the two manifolds can actually be related to each other by an $Sp(2n)$ transformation in a way similar to the conformal flatness of the conventional Minkowski and AdS space. This property will be discussed in detail in the next Section.

The $Sp(n)$ group element ${\cal O}(X)$, parametrized by the coordinates $X^{\mu\nu}$, defines Cartan forms $\Omega^{\alpha \beta}(X)$
\begin{equation}\label{Om}
{\cal O}^{-1} d {\cal O}= \Omega^{\alpha \beta} M_{\alpha \beta}=dX^{\mu\nu} E^{\alpha\beta}_{\mu\nu}(X)\,M_{\alpha \beta}\,.
\end{equation}
The Cartan forms
encode the vielbeine and the spin connections which characterize a geometry of $Sp(n)$. In eq. \eqref{Om} we distinguish the flat tangent--space basis on $Sp(n)$, labeled by the letters $\alpha,\beta, ...$ (from the beginning of the Greek alphabet), from the curved world basis associated with $X^{\mu\nu}$, labeled by the letters $\mu,\nu,...$ (from the middle of the Greek alphabet).

By construction, the Cartan forms \eqref{Om}
obey the Maurer--Cartan equations, which according to the algebra (\ref{algebrasp8})
have the form
\begin{equation}\label{MC}
d \Omega^{\alpha \beta} + \frac{\xi}{2} \Omega^{\alpha \gamma} \wedge \Omega_\gamma{}^\beta =0\,,
\end{equation}
where the indices are lowered and raised by $C_{\alpha\beta}$ and $C^{\alpha\beta}$ as in eq. \eqref{updown}.

As in the general case of the group manifolds, one can define a geometry of $Sp(n)$ to be flat with non-trivial torsion, or to have zero torsion and constant curvature.

In the zero--curvature geometry one chooses the spin connection to be zero and a local tangent--space basis to be formed by the vielbeine $E^{\alpha\beta}\equiv\Omega^{\alpha\beta}=dX^{\mu\nu} E^{\alpha\beta}_{\mu\nu}(X)$. From the Maurer--Cartan equations it then follows that the $Sp(n)$ torsion is
\be\label{T}
T^{\alpha\beta}=dE^{\alpha\beta}=-\frac{\xi}{2} E^{\alpha \gamma} \wedge E_\gamma{}^\beta\,.
\ee
The covariant derivatives associated with this geometry are constructed with the use of the inverse vielbeine
\be\label{nab}
\nabla_{\alpha\beta}=E_{\alpha\beta}^{\mu\nu}(X)\partial_{\mu\nu}\,,\qquad E_{\alpha\beta}^{\mu\nu}E_{\mu\nu}^{\gamma\delta}=\frac 12 (\delta_{\alpha}^\gamma\delta_\beta^\delta+\delta_{\alpha}^\delta\delta_\beta^\gamma)\,,
\ee
and form the  $Sp(n)$--algebra
\begin{equation} \label{algebra}
[\nabla_{\alpha \beta}, \nabla_{\gamma \delta}] = \frac{\xi}{4}(C_{\alpha \gamma} \nabla_{\beta \delta}
+ C_{\alpha \delta} \nabla_{\beta \gamma}
+ C_{\beta \gamma} \nabla_{\alpha \delta}
+ C_{\beta \delta} \nabla_{\alpha \gamma})\,.
\end{equation}
On the other hand, one can interpret \eqref{MC} as the torsion--free condition for the $Sp(n)$ geometry  with curvature, described by the vielbein $E^{\alpha\beta}$ and the connection $\omega_{\alpha}{}^{\beta}$ defined as follows
\be\label{Eom}
E^{\alpha\beta}=\Omega^{\alpha\beta}\,,\qquad \omega_{\alpha}{}^{\beta}=\frac\xi 4\Omega_{\alpha}{}^{\beta}\,.
\ee
The zero--torsion condition takes the form
\be\label{T0}
T^{\alpha\beta}=DE^{\alpha\beta}=dE^{\alpha\beta}+E^{\alpha\gamma}\wedge \omega_{\gamma}{}^{\beta}+E^{\beta\gamma}\wedge \omega_{\gamma}{}^{\alpha}=0\,,
\ee
and the $Sp(n)$ curvature is
\be\label{R}
R_\alpha{}^{\beta}=d\omega_\alpha{}^{\beta}+\omega_{\alpha}{}^\gamma\wedge \omega_{\gamma}{}^\beta=-\frac{\xi^2}{16}E_{\alpha}{}^\gamma\wedge E_{\gamma}{}^\beta\,.
\ee
The covariant differential
\be\label{Di}
D=E^{\alpha\beta}D_{\alpha\beta}=E^{\alpha\beta}\nabla_{\alpha\beta}+\omega\,,
\ee
acts on the contravariant and covariant spinors $F^\alpha$ and $F_\alpha$ as follows
\be\label{Dfup}
DF^\gamma=dF^\gamma+ F^\gamma\omega_{\gamma}{}^{\alpha}=E^{\alpha\beta}(\nabla_{\alpha\beta}\,\delta^\gamma_\delta+\frac{\xi} 8C_{\alpha\delta}\delta_\beta{}^\gamma
+\frac{\xi} 8C_{\beta\delta}\delta_\alpha{}^\gamma)F^\delta\,,
\ee
\be\label{Dfdown}
 DF_\gamma=dF_\gamma- \omega_{\alpha}{}^{\gamma}F_\gamma\,=E^{\alpha\beta}(\nabla_{\alpha\beta}\,\delta^\delta_\gamma-\frac{\xi} 8C_{\alpha\gamma}\delta_\beta{}^\delta
-\frac{\xi} 8C_{\beta\gamma}\delta_\alpha{}^\delta)F_\delta\,.
\ee
With the use of the zero--curvature covariant derivatives \eqref{nab}, the equations of motion of a bosonic field $B(x)$
 and a fermionic field  $F_\alpha(X)$, generalizing the flat hyperspace field equations \eqref{BF} and \eqref{FF} to  the $Sp(n)$ group manifold, have the following form  \cite{Plyushchay:2003tj}
\begin{eqnarray} \label{BADS}
&&(\nabla_{\alpha \beta} \nabla_{\gamma \delta}- \nabla_{\alpha \gamma} \nabla_{\beta \delta})B- \\ \nonumber
&&-\frac{\xi}{8}(C_{\alpha \gamma} \nabla_{\beta \delta}  - C_{\alpha \beta } \nabla_{\gamma \delta} + C_{\beta \delta} \nabla_{\alpha \gamma}- C_{\gamma \delta}\nabla_{\alpha \beta} + 2 C_{\beta \gamma} \nabla_{\alpha \delta} ) B-\\ \nonumber
&&  -(\tfrac{\xi}{8})^2 (C_{\alpha\gamma}C_{\beta\delta} - C_{\alpha \beta}C_{\gamma \delta}  + 2C_{\beta\gamma}C_{\alpha\delta})
B=0\,,
\end{eqnarray}
\begin{equation}\label{FADS}
\nabla_{\alpha \beta} F_\gamma - \nabla_{\alpha \gamma} F_\beta +\frac{\xi}{8}
   (C_{ \gamma \alpha} F_\beta - C_{ \beta \alpha} F_\gamma + 2 C_{\gamma \beta} F_\alpha    )=0\,.
 \end{equation}
 In the basis of the zero--torsion covariant derivatives \eqref{Dfdown}, the fermionic equation simplifies to
 \be\label{FADS1}
D_{\alpha \beta} F_\gamma - D_{\alpha \gamma} F_\beta=0\,,
\ee
while the bosonic equation takes the form
\be\label{BADS1}
(D_{\alpha \beta} D_{\gamma \delta}-D_{\alpha\gamma}D_{\beta\delta})B -(\tfrac{\xi}{8})^2 (C_{\alpha\gamma}C_{\beta\delta} - C_{\alpha \beta}C_{\gamma \delta}  + 2C_{\beta\gamma}C_{\alpha\delta})B=0\,.
\ee
Equations \eqref{BADS}--\eqref{BADS1} are $Sp(2n)$--invariant. This fact stems from the origin of these equations from the quantization of a corresponding $Sp(2n)$--invariant particle model \cite{Plyushchay:2003gv,Plyushchay:2003tj}. We will explicitly show this below using the generalized conformal flatness of the $Sp(n)$ manifold.

As we have mentioned, the flat hyperspace and the $Sp(n)$ group manifold can be realized as different cosets of their generalized conformal group $Sp(2n)$. This prompts one to ask whether their geometries, as well as the solutions of the scalar and spinor field equations in flat and $Sp(n)$ hyperspace, can locally be related by a generalized conformal transformation in a way similar to the conformally flat cases of conventional Minkowski and AdS spaces.

The answer to this question turns out to be positive. In order to explicitly demonstrate the connection between the two systems we will explore a special property of the $Sp(n)$ group manifolds found in \cite{Plyushchay:2003gv} and called `GL--flatness', where `GL' stands for `general linear' or `generalized conformal' flatness.

\section{GL--flatness of $Sp(n)$ group manifolds and the relation between the field equations in flat and $Sp(n)$ hyperspaces} \label{secgl4} \setcounter{equation}0

By GL--flatness of the $Sp(n)$ manifold we mean that, in a local coordinate basis associated with $X^{\alpha\beta}$, the covariant derivatives $\nabla_{\alpha\beta}$ \eqref{nab} satisfying the $Sp(n)$ algebra \eqref{algebra} take a very simple form
\begin{equation}\label{nabla}
\nabla_{\alpha \beta} = G_{\alpha}^{-1 \mu}(X) G_{\beta}^{-1 \nu}(X) \partial_{\mu \nu}\,,
\end{equation}
where $G_{\alpha}^{-1 \mu}(X)$ is a matrix which depends linearly on $X_{\alpha}{}^{\mu}$
\begin{equation}\label{sol2}
G_\alpha^{-1 \mu}(X) = \delta_\alpha^\mu +\frac{\xi}{4}X_\alpha{}^\mu\,.
\end{equation}
The corresponding $Sp(n)$
Cartan forms $\Omega^{\alpha \beta}$
are
\begin{equation}\label{sol1}
\Omega^{\alpha \beta} = d X ^{\mu \nu} G_{\mu}{}^\alpha(X) G_{\nu}{}^\beta(X)\,,
\end{equation}
where the matrix $G_{\mu}{}^\alpha(X) $ is inverse of $G_\alpha^{-1 \mu}(X)$ \footnote{Here we follow the notation of \cite{Plyushchay:2003gv} in which the matrix $G_{\beta}{}^\alpha(X) $ was introduced first and then $G_\alpha^{-1 \mu}(X)$ was derived as its inverse.} and has the following form
\begin{equation} \label{sol3}
G_\mu{}^{ \alpha}(X) = \delta^\alpha_\mu + \sum_{k=1}^\infty { \left( -\frac{\xi}{4} \right) }^k (X^k)_\mu{}^\alpha\,,
\end{equation}
where $(X^k)_\mu{}^\alpha$ stands for the product of the $k$ matrices $X_{\mu}{}^{\alpha}$.
Note that the possibility of representing the Cartan forms in the form \eqref{sol1} is a particular feature of the $Sp(n)$ group manifold since, in general, it is not possible to decompose the components of the Cartan form into a ``direct product" of components of some matrix  $G_{\mu}{}^\alpha$.

GL--flatness implies that the $Sp(n)$  Cartan forms  and the covariant derivatives can be obtained from the flat hyperspace ones by a transformation in the group $GL(n)\subset Sp(2n)$  involving the matrix $G_{\mu}{}^\alpha$ and its inverse.

The matrices $G_\alpha^{-1 \mu}(X)$ and $G_{\mu}{}^\alpha(X) $ satisfy the following identities
\be\label{100}
\partial_{\mu\nu}G^{-1\alpha\beta}=\frac\xi 8(\delta_\mu^\alpha\delta^\beta_\nu+\delta_\mu^\beta\delta^\alpha_\nu)\,,
\ee
\begin{equation}\label{101}
\partial_{\mu \nu} G_\rho{}^\sigma = \frac{\xi}{8}(G_{\rho \mu}G_\nu{}^\sigma + G_{\rho \nu} G_\mu{}^\sigma )\,,
\end{equation}
\begin{equation}\label{1011}
d G_\rho{}^\sigma = \frac{\xi}{4}  (\Omega_{\rho}{}^{\sigma}+2 G_{\rho}{}^\mu\Omega_\mu{}^\sigma)\,,
\end{equation}
\begin{equation}\label{104}
\partial_{\mu \nu} \sqrt {\det G} = \frac{\xi}{16} \sqrt{det G}\,(G_{\mu \nu} + G_{\nu \mu})\,,
\end{equation}
\begin{equation}\label{pG-1/2}
\partial_{\mu \nu} ({\det G})^{-\frac{1}{2}}= -\frac{\xi}{16} ({\det G})^{-\frac{1}{2}} (G_{\mu \nu} + G_{\nu \mu})\,,
\end{equation}
\begin{equation}\label{pG-D}
\partial_{\mu \nu} ({\det G})^{-\Delta}= -\frac{\xi\Delta}{8} ({\det G})^{-\Delta} (G_{\mu \nu} + G_{\nu \mu})\,,
\end{equation}
\begin{equation}\label{103}
	G_{\alpha\lambda}(X){G_\beta}^{\lambda}(X) = G_{[\alpha\beta]}(X)\equiv \frac 12(G_{\alpha\beta}(X)-G_{\beta\alpha}(X))\,.
\end{equation}
These identities can be used to check that equation (\ref{MC}) is indeed solved
by  (\ref{sol1})--(\ref{sol3}) and that the fields $B(X)$ and $F_\alpha(X)$ satisfying equations \eqref{BADS}--\eqref{BADS1} are related to the fields $b(X)$ and $f_\mu(X)$ satisfying the flat hyperspace equations \eqref{BF}--\eqref{FF} as follows
\be\label{defB}
B(X)=(\det G)^{-\frac 12}\,b(X)\,,
\ee
\be\label{fF}
 F_\alpha(X) =(\det G)^{-\frac 12}\, G^{-1}_\alpha{}^\mu f_\mu(X).
\ee
These relations are similar to the relations between the conformally invariant scalar and spinor equations in the conventional flat and $AdS$ spaces and reduce to them in the case of $n=2$, $D=3$.

\subsection{$Sp(2n)$ transformations of the fields on $Sp(n)$}
Using relations (\ref{defB}), (\ref{fF}), and the $Sp(2n)$ transformations \eqref{sp8fb}--\eqref{sp8ffD} of the bosonic and fermionic fields in flat hyperspace,  it is straightforward to derive the $Sp(2n)$ transformations of the fields on $Sp(2n)$.

Using the relation between the fields of weight $\Delta= \frac{1}{2}$ on flat hyperspace and on $Sp(n)$ group manifold  \eqref{defB}
we have the following relation between the $Sp(2n)$ transformations of the wight--$\frac{1}{2}$  fields on $Sp(n)$ and in flat hyperspace
\if{}
\be\label{DB}
B'(X')=({\det G(X')})^{-\frac 12}\,b'(x')\,\quad F'_\alpha(X')=({\det G(X')})^{-\frac 12} \,G_\alpha^{-1\mu}(X')\, f'_{\mu}(X')
\ee
\fi
\be\label{db}
\delta B(X)=({\det G})^{-\frac 12}\delta b\,,
\ee
\be\label{dfF}
\delta F_\alpha=({\det G})^{-\frac 12} \,G_\alpha^{-1\mu}\,\delta f_{\mu}\,.
\ee
Note that in the above expressions the matrix $G_\alpha{}^\mu$ is not varied since it is form--invariant, i.e. $G(X')$ has the same form as $G(X)$.

Then, in view of eq. \eqref{pG-1/2} the $Sp(n)$--variations of $B(X)$ and $F_\alpha(X)$ have the following form
\begin{equation}\label{sp8fB-01}
\delta {B}(X)= -(a^{\alpha \beta} {\cal D}_{\alpha \beta} +\frac 12  (g_\alpha{}^\alpha- k_{\alpha \beta}X^{\alpha \beta})+ 2 g_\beta{}^\alpha X^{\beta \gamma}
{\cal D}_{\alpha \gamma} -
 k_{\alpha \beta}  X^{\alpha \gamma} X^{\beta \delta}{\cal D}_{\gamma \delta}) {B}(X)\,,
\end{equation}
\begin{eqnarray}\label{sp8fF-01}
&\delta {F_\sigma}(X)= -(a^{\alpha \beta} {\cal D}_{\alpha \beta} + \frac 12  (g_\alpha{}^\alpha-k_{\alpha \beta} X^{\alpha \beta}) + 2 g_\beta{}^\alpha X^{\beta \gamma}
{\cal D}_{\alpha \gamma} -
 k_{\alpha \beta}  X^{\alpha \gamma} X^{\beta \delta}{\cal D}_{\gamma \delta}) {F_\sigma}(X)\,,&\nonumber\\
 &-(g_\sigma{}^\beta - k_{\sigma \alpha}X^{\alpha \beta}) F_\beta(X)\,,&
\end{eqnarray}
where the derivative ${\cal D}_{\alpha \beta}$ is defined as
\begin{equation}\label{cD0}
{\cal D}_{\alpha \beta}= \partial_{\alpha \beta} + \frac{\xi}{16} (G_{\alpha \beta} + G_{\beta \alpha})\,.
\end{equation}
Using  (\ref{101}) one can check that these derivatives commute with each other
$[ {\cal D}_{\alpha \beta}, {\cal D}_{\gamma \delta} ]=0$ just as in the flat case.

Let us note  that the relation between the flat and $Sp(n)$ hyperfields of an arbitrary weight $\Delta$ and the form of the corresponding $Sp(2n)$ transformations require additional study since to this end one should know the form of $Sp(2n)$--invariant equations satisfied by these fields. In this respect, the results of \cite{Gelfond:2003vh,Gelfond:2013lba} on higher--rank hyperfields and currents can be useful. This issue will be addressed elsewhere.

\section{$AdS_4$ Metric}\label{secads} \setcounter{equation}0

Before considering correlation functions, let us first  demonstrate the connection between the $Sp(4)$ group manifold  and
$AdS_4$ space explicitly in the GL--flat basis \eqref{sol1}, \eqref{sol3}. In order to do so,
we shall  compute  an explicit form of the $x^m$--dependent part of the  metric on the $Sp(4)$ group manifold in the $GL(4)$
flat parametrization and prove that it corresponds to a specific parametrization of the $AdS_4$ metric.
In other words,  we have to evaluate the expression
\begin{equation} \label{EX}
\Omega^{\alpha \beta}(x^m) =\frac{1}{2} dx^m {(\gamma_m)}^{\delta \sigma}G_\delta{}^{\alpha} G_{\sigma}{}^\beta =
\frac{1}{2} dx^m e_m^a {(\gamma_a)}^{\alpha \beta} + \frac{1}{4} dx^m \omega_m^{ab} {(\gamma_{ab})}^{\alpha \beta},
\end{equation}
 where the dependence of the matrices $X^{\alpha \beta}$ on the coordinates $y^{mn}$  (see eq. \eqref{X}) is discarded, i.e. $X_\alpha{}^\beta = \frac{1}{2}x^n (\gamma_n)_\alpha{}^\beta$. Denoting $x^2 = x^m x^n \eta_{mn}$ and  $x_m = \eta_{mn} x^n$ and, using the explicit form (\ref{sol3}) of $G_\mu{}^\alpha(X)$, one obtains
\begin{align}
	\Omega^{\alpha\beta}(x)= \frac{1}{2} \frac{dx^m}{[1-(\frac{\xi}{8})^2 x^2]^2}\left[(\gamma_\ell)^{\alpha\beta}\left([1+(\tfrac{\xi}{8})^2 x^2]\delta_m^\ell-2(\tfrac{\xi}{8})^2 \eta_{mn}x^n x^\ell\right)-\tfrac{\xi}{4}x^n (\gamma_{mn})^{\alpha\beta}\right] \,.
\end{align}
Hence,  the vierbein and spin--connection take the form:
\begin{align}
	e^a_m &= \frac{1}{[1-(\frac{\xi}{8})^2 x^2]^2}\left( [1+(\tfrac{\xi}{8})^2 x^2]\delta_m^a-2(\tfrac{\xi}{8})^2
x^a x_m\right) \,,\\
	\omega^{ab}_{m} &= \frac{-2\xi}{[1-(\frac{\xi}{8})^2 x^2]^2}\,\delta^{[a}_m x^{b]} = - \frac{8 (\frac{\xi}{8})}{(1-(\frac{\xi}{8})^2 x^2)^2}(x^a \delta^b_m-x^b \delta^a_m)\,.
\end{align}
For  completeness, let us also present   the explicit form of the metric, the inverse vierbein and the inverse metric
\begin{align}\label{OurMetric}
	g_{mn}= \frac{1}{[1-(\frac{\xi}{8})^2 x^2]^4}\left( [1+(\tfrac{\xi}{8})^2 x^2]^2 \eta_{mn}-4 (\tfrac{\xi}{8})^2 x_m x_n \right) \,,
\end{align}
\begin{align}
	e^m_a = \frac{1-(\frac{\xi}{8})^2 x^2}{1+(\frac{\xi}{8})^2 x^2} \left( [1-(\tfrac{\xi}{8})^2 x^2]\delta^m_a + 2(\tfrac{\xi}{8})^2 x_a x^m\right)~,
\end{align}
\begin{align}
	g^{mn}= \frac{[{1-(\frac{\xi}{8})^2 x^2]}^2}{{[1+(\frac{\xi}{8})^2 x^2]}^2}\left( [1-(\tfrac{\xi}{8})^2 x^2]^2   \eta^{mn} + 4 (\tfrac{\xi}{8})^2 x^m x^n \right) \,.
\end{align}
It is well-known that the $AdS_D$ metric (\ref{OurMetric}) can be represented as an embedding in a flat $(D+1)$-dimensional
space
\begin{align}\label{embedmetric}
	ds^2 = \eta_{mn} dy^m dy^n - (dy^D)^2~,
\end{align}
via the embedding constraint
\begin{align}
	\eta_{mn}y^m y^n - (y^D)^2 = -r^2~.
\end{align}
Choosing the embedding coordinates for $AdS_4$ to be
\begin{align}\label{Fronsdal2}
	& y^m = \frac{1+(\frac{\xi}{8})^2 x^2}{[1-(\frac{\xi}{8})^2 x^2]^2}\, x^m, \qquad
 y^4 = \sqrt{r^2 + x^2 \frac{1+(\frac{\xi}{8})^2 x^2}{[1-(\frac{\xi}{8})^2 x^2]^2}}~,
\end{align}
one readily recovers the metric  (\ref{OurMetric}), with the parameter $\xi$ being related to the $AdS_4$ radius $r$ through
\begin{align}\label{relationxi}
	\xi  = \frac{2}{r}~.
\end{align}
Finally, computing the Riemann tensor
\begin{align}
	{R^{ab}}_{mn}= -32(\tfrac{\xi}{8})^2 \frac{1+(\frac{\xi}{8})^2x^2}{[1-(\frac{\xi}{8})^2x^2]^4}\Bigr( [1+
( \tfrac{\xi}{8})^2x^2 ]\delta^{[a}_{m}\delta^{b]}_n + 4(\tfrac{\xi}{8})^2 x^{[a}\delta^{b]}_{[m}x_{n]} \Bigr)~,
\end{align}
and  the Ricci scalar
\begin{align}
	R = - 192 \left(\frac{\xi}{8}\right)^2 = - 3\xi^2~,
\end{align}
one verifies that the metric   (\ref{OurMetric}) indeed  corresponds to a  space with constant negative curvature.
\if{}
\textbf{In a similar way one may derive the form of the vielbeine and spin connection of the AdS--space of an appropriate dimension $D$ as a subspace of $Sp(n)$. For instance, $AdS_6$ in the $n=8$ case and $AdS_{10}$ in $n=16$.}
\fi

We are now in a position to consider $Sp(2n)$--invariant correlation functions of the hyperfields.

\section{Correlation functions on $Sp(n)$ group manifold} \setcounter{equation}0 \label{seccf}

One can derive the generic form of  $Sp(2n)$--invariant correlation functions for bosonic and fermionic fields of weight--$\frac 12$ on
the $Sp(n)$ group manifolds in a way similar to the conventional conformal field
theories in various dimensions \cite{Osborn:1993cr} (see also \cite {Polyakov:1970xd} for analogous computations in two--dimensional CFTs), as was carried out in \cite{Vasiliev:2003jc} for computing the two-- and three--point correlation functions in flat hyperspace. Because of the GL--flatness property of the hyperspaces, the correlation functions are related by the generalized conformal transformation.

For instance, since the two--point correlation functions for the flat--space fields $b$ and $f_\mu$ of conformal weight--$\frac 12$  satisfy the free equations, they are related to the corresponding two--point functions in $Sp(n)$ in the same way as the fields themselves, i.e
\begin{equation}\label{2pb}
 \langle B(X_1) B(X_2)\rangle_{Sp(n)}= c_B\,( {\det G (X_1 ))}^{-\frac 12}  ( {\det G (X_2 ))}^{-\frac 12}       (\det |X_{12}|)^{- \frac 12}\,,
\end{equation}
\bea\label{2pf}
\langle F_\alpha (X_1) F_\beta(X_2)\rangle_{Sp(n)}=\hspace{275pt}&\\
 c_F\,  G_\alpha^{-1\mu}(X_1)\, G_\beta^{-1\nu}(X_2)\,    ( {\det G (X_1 ))}^{-\frac 12}  \,( {\det G (X_2 ))}^{-\frac 12}
  (X_{12})^{- 1}_{\mu\nu}\,(\det |X_{12}|)^{- \frac 12}\,,& \nonumber
\eea
where $G(X_1)$ and $G(X_2)$ stand, respectively, for  $G_{\alpha}{}^{\beta}(X_1)$ and $G_{\alpha}{}^{\beta}(X_2)$,
$c_B$ and $c_F$ are constants which are not fixed by the $Sp(2n)$ invariance, $X_{ij}= X_i -X_j$ and
$$
\langle b(X_1) b(X_2)\rangle_{flat}= c_b(\det |X_{12}|)^{- \frac 12}\,,
 $$
 $$
 \langle f_\mu (X_1) f_\nu(X_2)\rangle_{flat}=c_f (X_{12})^{-1}_{\mu\nu}\,(\det |X_{12}|)^{- \frac 12}
$$
are the flat space correlation functions computed in \cite{Vasiliev:2001dc}.

\if{}
In the same way, one may relate the flat hyperspace and $Sp(n)$ $N$--point functions of the fields of arbitrary conformal weight $\Delta$, or to perform the direct derivation of the $Sp(n)$ $N$--point functions by requiring their invariance under the $Sp(2n)$ transformations \eqref{sp8fB-01}, as we shall do below.
\fi

Let us comment on the conformal dimensions of the various fields entering the correlation functions. For a flat hyperspace,
the hyperfields have conformal weight $\Delta=\frac{1}{2}$ and are identified with the primary fields of the conformal field theory, whereas conformal fields with higher conformal weights correspond to
derivatives of the  hyperfields and are identified with the descendants. One may also consider primary fields of higher weight in hyperspace \cite{Gelfond:2003vh,Gelfond:2013lba} which are products of the master fields. For example, a bilinear combination of master fields corresponds to
conserved currents and  they are dual to master fields in higher dimensional hyperspaces. The detailed study of the
generalized Conformal Field Theory involving such composite operators will be addressed in future work.
Hence, in what follows, when relating correlation functions on flat and AdS hyperspaces, we will assume the fields to have weight $\frac{1}{2}$.

\subsection{Two--point functions}
Let us denote by
\begin{equation}
{\Phi}(X_1, X_2) = \langle B(X_1) B(X_2)  \rangle_{Sp(n)} \,,
\end{equation}
the two--point correlation function of two scalar fields of conformal weight $\frac 12$  on $Sp(n)$.
The invariance under the transformations \eqref{sp8fB-01} generated by the parameter $a^{\alpha \beta}$  results in the equation
\begin{equation}\label{2pa}
a^{\alpha \beta}({ \cal D}_{1, \alpha \beta} + {\cal D}_{2, \alpha \beta} )\Phi(X_1, X_2)=0 \,.
\end{equation}
In view of the identity
\begin{equation}
\frac{\partial}{ \partial X^{\alpha \beta}} \det |X| = X_{\alpha \beta}^{-1} \det |X|, \qquad
X^{\alpha \gamma} X_{\gamma \beta}^{-1} = \delta^\alpha_\beta \,,
\end{equation}
equation  (\ref{2pa})  is solved by
\begin{equation}
\Phi(X_1, X_2) = ( {\det G (X_1 ))}^{-\frac{1}{2}}  ( {\det G (X_2 ))}^{-\frac{1}{2}} \,{\tilde \Phi}(\det |X_{12}|)\,,
\end{equation}
where here,  $ {\tilde \Phi}(\det |X_{12}|)$ is an arbitrary function of   $\det |X_1 - X_2|$.
Imposing also the invariance of the two-point function under the transformations
generated by the parameter $g_\alpha{}^\beta$, namely
\begin{equation}
(g_\alpha{}^{\alpha} + 2 g_\beta{}^\alpha( X_1^{\beta \gamma} {\cal D}_{1,\alpha \gamma}+
 X_2^{\beta \gamma} {\cal D}_{2,\alpha \gamma}) )\Phi(X_1, X_2)=0\,,
\end{equation}
fixes the form of the function  ${\tilde \Phi}(\det |X_1 - X_2|)$ and results in the following expression for the two--point function
\begin{equation}\label{2p}
 \Phi(X_1, X_2)=  c_B     ( {\det G (X_1 ))}^{-\frac{1}{2}}  ( {\det G (X_2 ))}^{-\frac{1}{2}}
  (\det |X_{12}|)^{-\frac{1}{2}}\,,
\end{equation}
where $c_B$ is an arbitrary constant.
Finally, the invariance under the transformations generated by the parameters $k^{\alpha \beta}$
imposes the condition
\begin{equation}
k_{\alpha \beta} (\tfrac{1}{2} X_1^{\alpha \beta} + X_1^{\alpha \gamma} X_1^{\beta \delta} {\cal D}_{1,\gamma \delta}
+
\tfrac{1}{2} X_2^{\alpha \beta} + X_2^{\alpha \gamma} X_2^{\beta \delta} {\cal D}_{2,\gamma \delta})\,
 \Phi(X_1, X_2)=0\,,
\end{equation}
which is identically satisfied by \eqref{2p}.

The derivation of eq. \eqref{2p} reproduces the relation \eqref{2pb} between the $Sp(n)$ two--point functions of scalar fields of conformal weight $\frac 12$ with those in flat hyperspace computed in \cite{Vasiliev:2003jc}. Analogously, one may check the relation \eqref{2pf} between the two--point functions of two spinor fields of weight $\frac 12$.

\subsection{Three--point functions}

The calculation of the weight--$\frac 12$ bosonic field three--point function
\begin{equation}
\Phi(X_1, X_2, X_3) = \langle B(X_1) B(X_2)     B(X_3) \rangle_{Sp(n)}\,,
\end{equation}
proceeds in a similar way. From the equation
\begin{equation}
a^{\alpha \beta}({ \cal D}_{1,\alpha \beta} + {\cal D}_{2,\alpha \beta} + {\cal D}_{3,\alpha \beta} )\Phi(X_1, X_2,X_3)=0\,,
\end{equation}
one obtains
 \begin{eqnarray} \nonumber
\Phi(X_1, X_2,X_3)& =& c_3\,( {\det G (X_1 ))}^{-\frac{1}{2}} ( {\det G (X_2 ))}^{-\frac{1}{2}} ( {\det G (X_3 ))}^{-\frac{1}{2}}\\
 &&{\tilde \Phi}(\det|X_{12}|, \det |X_{23}|, \det |X_{13}|)\,,
\end{eqnarray}
where $\tilde \Phi$ is an arbitrary function  depending only on the combinations $\det |X_i - X_j| $.
The equation
\begin{equation}
\Bigr(\tfrac{3}{2} g_\alpha{}^{\alpha} + 2 g_\beta{}^\alpha \sum_{i=1}^{i=3} X_i^{\beta \gamma} {\cal D}_{i,\alpha \gamma}
 \Bigr)F(X_1, X_2,X_3)=0\,,
\end{equation}
then fixes the form of the function $\Phi$ to be
\begin{eqnarray} \label{3pb}\nonumber
\Phi(X_1, X_2,X_3)&=&   ( {\det G (X_1 ))}^{-\frac{1}{2}} ( {\det G (X_2 ))}^{-\frac{1}{2}} ( {\det G (X_3 ))}^{-\frac{1}{2}}
\\
&&\!\!\! \!\!\!\!\!\! {(\det |X_{12}|)}^{-\frac{k_3}{2}} \,{(\det |X_{23}|)}^{-\frac{k_1}{2}}\,{(\det |X_{13}|)}^{-\frac{k_2}{2}}\,,
\end{eqnarray}
with
\begin{equation}
k_1+k_2+k_3= \frac 32 \,.
\end{equation}
Finally,  the equation
\begin{equation}
k_{\alpha \beta} \sum_{i=1}^{i=3}(\frac 12 X_i^{\alpha \beta} + X_i^{\alpha \gamma} X_i^{\beta \delta}
{\cal D}_{i , \gamma \delta})\,\Phi(X_1, X_2,X_3)=0\,,
\end{equation}
implies that
\begin{equation}
k_1=k_2=k_3=\frac 12.
\end{equation}
The equation \eqref{3pb} relates (via the conformal factors $ ({\det G (X_1 ))}^{-\frac{1}{2}} ( {\det G (X_2 ))}^{-\frac{1}{2}} ( {\det G (X_3 ))}^{-\frac{1}{2}} $) the bosonic field three--point function on $Sp(n)$ with that in flat hyperspace computed in \cite{Vasiliev:2003jc}. In a similar way, one observes that the three--point function involving two spinor and one scalar fields\footnote{The correlation functions containing an odd number of spinor fields vanish identically \cite{Vasiliev:2003jc}.} are related as follows
\bea\label{3pf}
&\!\!\!\!\!\!\!\!\!\!\!\!\langle F_\alpha(X_1) F_\beta(X_2) B(X_3)\rangle_{Sp(n)}=
c_{3f} \, ( {\det G (X_1 ))}^{-\frac{1}{2}} ( {\det G (X_2 ))}^{-\frac{1}{2}} ( {\det G (X_3 ))}^{-\frac{1}{2}}&
\\
&G^{-1\mu}_{\alpha}(X_1)\,G^{-1\nu}_{\beta}(X_2)\,(X_{12})^{-1}_{\mu\nu}\,{(\det |X_{12}|)}^{-\frac 14}\,{(\det X_{23}|)}^{-\frac 14} \,{(\det |X_{13}|)}^{-\frac 14} \,.&\nonumber
\eea

\subsection{Four--point functions}
To the best of our knowledge, the explicit form of $N$--point functions for $N\geq 4$, both in flat and $Sp(n)$ hyperspace,
 have not previously  been given in the literature, and below we present the result for the four--point correlation functions \footnote{In the unfolded formulation of hyperspace dynamics, $N$--point correlation functions  were computed in \cite{Didenko:2012tv}, however their relation to our form of the correlators still remains to be understood.}.

The computation of the $Sp(2n)$--invariant four--point functions follows the same lines as the computation of the two-- and three--point functions.
Let us consider first the correlation function of four scalar fields of an arbitrary weight $\Delta$ in flat hyperspace. Its invariance under the translations
\begin{equation}
a^{\alpha \beta}\sum_{i=1}^4\frac{\partial }{ \partial X_i^{\alpha \beta}}  \Phi(X_1, X_2,X_3,X_4)=0\,,
\end{equation}
 implies  that the function $\Phi(X_1,X_2,X_3, X_4)$ depends only on the differences
$X_{ij} = X_i - X_j$. Using the analogy with the usual conformal field theory for a four--point function  we write
\begin{equation}\label{4PF}
\Phi(X_1, X_2, X_3, X_4) = c_4\,\prod_{ij, i<j}\frac{1}{{(\det |X_{ij}|)}^{\Gamma_{ij}}   }
{\tilde \Phi} \left (  z,z' \right)\,,
\end{equation}
where $z,z'$ are the two independent cross-ratios
\begin{equation}
	z= \det\left(\frac{|X_{12}||X_{34}|}{|X_{13}||X_{24}|}\right)~,~z'= \det\left(\frac{|X_{12}||X_{34}|}{|X_{23}||X_{14}|}\right)\,.
\end{equation}
Crossing symmetry then implies the constraint
\begin{equation}\label{crossingSym}
	 \tilde\Phi(z,z')=\tilde\Phi\left(\frac{1}{z},\frac{z'}{z}\right)=\tilde\Phi\left(\frac{z}{z'},\frac{1}{z'}\right)\,.
\end{equation}
Then, requiring the invariance under the $GL(n)$ transformations, namely
\begin{equation}
\left( g_\alpha{}^{\alpha}\sum_{i=1}^{i=4}\Delta_i  + 2 g_\beta{}^\alpha \sum_{i=1}^{i=4} X_i^{\beta \gamma}
\frac{\partial}{\partial X^{\alpha \beta}_i} \right)
 \Phi(X_1, X_2,X_3, X_4)=0 \,,
\end{equation}
one obtains  the additional condition
\begin{equation}
\sum_{j=2}^{j=4}\,\,\, \sum_{i =1}^{i=j-1} \Gamma_{ij} = \frac{1}{2} \sum_{i=1}^{i=4} \Delta_i \,.
\end{equation}
Finally, the invariance under the generalized conformal boosts
\begin{equation}
k_{\alpha \beta} \sum_{i=1}^{i=4}\left( \Delta_i X_i^{\alpha \beta} + X_i^{\alpha \gamma} X_i^{\beta \delta}
\frac{\partial}{\partial X_i^{\gamma \delta} } \right) \Phi(X_1, X_2,X_3, X_4)=0\,,
\end{equation}
imposes the condition on the conformal weights
\begin{equation}
\sum_{j} \Gamma_{ij} =  \Delta_i, \quad i\neq j \,.
\end{equation}
Let us note that, similar to the usual conformal field theory, the four--point function
(\ref{4PF}) contains an arbitrary function $\tilde \Phi$, whose argument can be considered as a
generalization of the cross--ratios to the case of matrix--valued coordinates $X^{\mu\nu}$.

Now, as in the case of the two-- and three--point functions, the expressions for four--point functions on the  $Sp(n)$ group manifold for the primary weight--$\frac{1}{2}$ fields can be obtained from the corresponding expressions on the flat hyperspace by the re--scaling of the former with appropriate factors of ${(\det G (X_i))}^{-\frac{1}{2}}$, i.e.
\begin{equation}
\Phi(X_1, X_2, X_3,X_4)_{Sp(4)} = \Phi(X_1,X_2,X_3,X_4)\prod_{i=1}^4{(\det G (X_i))}^{-\frac{1}{2}}\,,
\end{equation}
where the function $\Phi(X_1,X_2,X_3,X_4)$  on the right hand side of this equation is given in (\ref{4PF}).

In the same way, one can obtain the four--point function of four spinor fields of weight $\Delta=\frac{1}{2}$ by rescaling the flat hyperspace result
\begin{equation}
	\begin{split}
	&\langle F_\alpha(X_1)F_\beta(X_2)F_\gamma(X_3)F_\delta(X_4)\rangle_{Sp(4)}\\
&={G^{-1}_\alpha}^\mu(X_1){G^{-1}_\beta}^\nu(X_2){G^{-1}_\gamma}^\rho(X_3){G^{-1}_\delta}^\sigma(X_4) \prod_{i=1}^4 (\det G(X_i))^{-\frac{1}{2}}\langle F_\mu(X_1)F_\nu(X_2)F_\rho(X_3)F_\sigma(X_4)\rangle_{flat}
\end{split}
\end{equation}
\bea\label{3fflat}
&\langle F_\mu(X_1)F_\nu(X_2)F_\rho(X_3)F_\sigma(X_4)\rangle_{flat}=
\prod_{i<j}\det|X_{ij}|^{-\frac{1}{3}}\big[(X_{12})^{-1}_{\mu\nu}(X_{34})^{-1}_{\rho\sigma}\Phi_{12,34}(z,z')\nonumber\\
&-(X_{13})^{-1}_{\mu\rho}(X_{24})^{-1}_{\nu\sigma}\Phi_{13,24}(z,z')
+(X_{14})^{-1}_{\mu\sigma}(X_{23})^{-1}_{\nu\rho}\Phi_{14,23}(z,z')\big]\,.
\eea

As before, the functions $\Phi_{ij,k\ell}(z,z')$ are indeterminate functions of the crossing ratios constrained by crossing symmetry to satisfy
\begin{equation} \Phi_{12,34}(z,z')=\Phi_{13,24}\left(\frac{1}{z},\frac{z'}{z}\right)=\Phi_{14,23}\left(\frac{z}{z'},\frac{1}{z'}\right)\,.
\end{equation}
Finally, one may obtain the four--point function that involves two spinorial and two bosonic fields on flat hyperspace
\be\label{42fb}
\langle F_\mu(X_1)F_\nu(X_2)b(X_3)b(X_4)\rangle_{flat}=(X_{12})_{\mu\nu}^{-1} \,\tilde\Psi(z,z') \,\prod_{i<j}\det|X_{ij}|^{-\frac{1}{3}}\,,
\ee
and relate it to the corresponding correlator on $Sp(n)$
\begin{equation}
	\begin{split}
&	\langle F_\alpha(X_1)F_\beta(X_2)b(X_3)b(X_4)\rangle_{Sp(4)}\\
&={G^{-1}_\alpha}^\mu(X_1) {G^{-1}_\beta}^\nu(X_2) \prod_{i=1}^4 (\det G(X_i))^{-\frac{1}{2}}\, \langle F_\mu(X_1)F_\nu(X_2)b(X_3)b(X_4)\rangle_{flat}\,.
	\end{split}
\end{equation}
In eq. \eqref{42fb}, again, $\Psi(z,z')$ is a function  satisfying the crossing relations \eqref{crossingSym} that cannot be determined by the $Sp(2n)$--symmetry alone. In fact, the functions of the cross--ratios $\tilde\Phi(z,z'), \Phi_{ik,k\ell}(z,z')$ and $\Psi(z,z')$
should be  completely determined in terms of the OPE between primary fields in the CFT on flat hyperspace.

\section{Conclusion} \setcounter{equation}0

In this paper we have considered some aspects of the $Sp(2n)$ invariant formulation
of higher spin fields. The main advantage of this approach is that one may combine infinite series of higher spin fields into one scalar and one  spinor ``master" field defined on  a hyperspace. It is then possible to
  study their field equations, correlation functions and other properties by making appropriate generalizations of analogous quantities for  scalar and spinor fields on conventional flat and AdS spaces.

We have considered the theory on both, the flat hyperspace and the $Sp(n)$ group manifold, the latter being a hyperspace extension of $AdS_d$  space.
As we mentioned above, it is quite instructive to follow the analogy  between the properties of conformal scalar and spinor fields on a flat space and
a scalar and a spinor field on anti de Sitter space on the one side, and a scalar and spinor field  on flat and $Sp(n)$ hyperspace on the other side.

Typically, computations on the ordinary AdS space are performed in a particular conformally flat parametrization of the metric. Similarly, our study of the field equations and correlation functions on $Sp(n)$ group manifolds
 has been heavily based on the  $GL(n)$--flatness property of these manifolds.
By exploiting this property, we have established, via the generalized conformal transformation, the relation between the field equations for scalar and spinor ``master" fields on flat and $Sp(n)$ group manifolds.

Provided that the fields have an appropriated conformal weight ($\Delta=\frac 12$), their equations of motion on flat hyperspace and $Sp(n)$ group manifold are invariant under the $Sp(2n)$ symmetry, which is a generalization of the usual conformal symmetry to the case of the hyperspaces.
 Using a technique similar to that of multidimensional conformal field theories \cite{Osborn:1993cr}, we have extended the results of \cite{Vasiliev:2001dc,Vasiliev:2003jc} on two-- and three--point functions of the scalar and spinor hyperfields on flat hyperspace by deriving the form of the two-- and three--point correlation functions on $Sp(n)$ and four--point functions on flat and $Sp(n)$ hyperspace exploiting their invariance under the $Sp(2n)$ group. The correlation functions on the $Sp(n)$ manifolds are related to their flat hyperspace counterparts by the the generalized conformal transformation similar to that related the fields themselves.

Let us note that the results obtained in this paper are valid for $Sp(2n)$ invariant theories with an arbitrary value of $n$. By now, however, the most studied physically interesting example has been the case of $n=4$
which corresponds to the four--dimensional higher spin gauge theory. In this case,
the master fields $b(X)$ and $f_\alpha(X)$ contain the curvatures of all the higher spin fields (with the spin ranging from zero to infinity) and the hyperfield equations of motion encode the Bianchi identities and the field equations for the higher spin curvatures explicitly derived in the flat space case only (see Section \ref{Hyperflat}). So it will be instructive to derive in a similar way the Bianchi identities and the equations of motion for the higher spin curvatures in $AdS_4$ from the field equations \eqref{BADS}--\eqref{BADS1} on $Sp(4)$. In this case it is not consistent anymore to naively extend the $Sp(4)$ hyperfields $B(x,y)$ and $F_\alpha(x,y)$ as series in the powers of the tensorial coordinates $y^{mn}$ (as in the flat case \eqref{ymn}), but one should rather perform the expansion in the Lorentz harmonics of the group $SO(3,1)\subset Sp(4)$ parametrized by $y^{mn}$, very much like when performing the series expansion of fields in Kaluza--Klein theories. To this end, it might be useful to choose a different parametrization of the $Sp(4)$ group manifold of the form ${\mathcal O}(x^l, y^{mn})=K(x)H(y)$, where $K(x)$ is the $AdS_4$ coset element and $H(y)$ is the $SO(3,1)$ group element.

It would also be of interest to apply the results of this paper to the description of higher spin theories in higher dimensions, especially in AdS. For instance, the cases of $n=8$ and $n=16$  correspond, respectively, to conformal higher spin fields on six-- and ten--dimensional space--time, which was demonstrated for theories in flat (hyper)space in \cite{Bandos:2005mb}, while the extension of these results to theories in AdS is still to be carried out.

We hope that the results obtained in this paper will be useful for better understanding generic conformal properties of higher spin fields (see \emph{e.g.} \cite{Metsaev:2008ks,Metsaev:2009ym,Metsaev:2013wza,Gover:2008sw},
 \cite{Costa:2011mg,Maldacena:2011jn,Stanev:2012nq}),   as well as for further study of higher spin AdS/CFT duality.

However, the most ambitious issue in this kind of theories is the interaction problem. As we have seen, the generalized conformal field theories in hyperspaces considered above have nontrivial three-- and four--point correlation functions, however this does not yet imply the existence of non--trivial interactions.  One should resort to additional criteria, such as the study of anomalous dimensions of conformal  operators  etc. See e.g. \cite{Petkou:1994ad} for the discussion of these issues in conventional CFTs. If non--trivial interactions of scalar and spinor hyperfields exist, from the perspective of higher spin field theory they should correspond to higher--order terms in higher spin curvatures contained inside the hyperfields. This is yet another important open problem of how to incorporate higher spin potentials directly into the hyperspace framework. So far this has only been done with the use of the unfolded technique \cite{Vasiliev:2007yc}.

So, it seems to be particularly interesting to perform a further detailed study of properties of conformal field theories on flat  hyperspaces and $Sp(n)$ group manifolds and, in particular, the possibility of constructing interacting conformal field theories on these spaces. In this respect, let us note that upon having verified the consistency associated with gauge invariance etc. (see \emph{e.g.} \cite{Fotopoulos:2008ka}-\cite{Tsulaia:2012rb},\cite{Boulanger:2013zza} and references therein),  consistency checks of interacting higher spin gauge theories on flat and AdS backgrounds proceed in different ways. For a flat background extra crucial constraints
on consistent interactions are obtained by requiring the existence of a nontrivial S--matrix (in contrast to the theories on AdS backgrounds in which the conventional notion of S--matrix is not applicable) and these constraints appear when one considers quartic interaction vertices \cite{Sagnotti:2010at,Fotopoulos:2010ay,Taronna:2011kt,Dempster:2012vw}.

\subsection*{\bf Acknowledgments}
We are grateful to P. Dempster, S. Kuzenko, A. Petkou, I. Samsonov, E. Skvortsov and M. Vasiliev for fruitful discussions. I.F. would like to acknowledge the Theory Division at CERN and the Max--Planck--Institut f\"ur Gravitationsphysik in Potsdam (Golm) for their warm hospitality during the later stages of this work.
 Work of D.S. was partially supported by the MIUR-PRIN contract 2009-KHZKRX and by the INFN Special Initiative TV12. D.S. acknowledges hospitality and support extended to him at CBPF, Rio de Janeiro and at ICTP SAIFR, Sao Paulo, Brazil during the Workshop ``Higher-Spin and Higher-Curvature Gravity" at the final stage of this project.
M.T. would like to thank the Department of Physics, the University of Auckland, where part of this work has
been performed, for its kind hospitality. Work of M.T.  has been supported in part by an  Australian Research Council  grant DP120101340. M.T. would also like to acknowledge grant   31/89 of the Rustaveli National Science Foundation.

\setcounter{equation}0
\appendix
\numberwithin{equation}{section}

\section{Proof of the $GL(n)$ flatness, some technical details and useful identities}\label{Appendix A}

We use the following normalization for the Dirac $\gamma$--matrices
\begin{equation}\label{1}
(\gamma^m)^\alpha{}_\delta (\gamma^n)^\delta{}_\beta
+
(\gamma^n)^\alpha{}_\delta (\gamma^m)^\delta{}_\beta
 = 2 \eta^{mn} \delta^\alpha_\beta \,,
\end{equation}
where $m$,  $n$ and other Latin letters are space-time vector indices,  and
$\alpha, \beta$ and other Greek letters correspond to spinorial indices.
Throughout the paper  ``$(,)$" denotes symmetrization and ``$[,]$" denotes antisymmetrization with weight one.
The symplectic matrix $C^{\alpha \beta}=-C^{\beta\alpha}$  is used
to relate upper and lower spinorial indexes as follows
\begin{equation}\label{updown}
\mu^{\alpha}= C^{\alpha \beta} \mu_\beta, \quad \mu_{\alpha}= -C_{\alpha \beta} \mu^\beta,
\quad C^{\alpha \gamma}C_{\gamma \beta}=-\delta^\alpha_\beta \,.
\end{equation}

We shall now show that the Cartan form given by  (\ref{sol1})--(\ref{sol3}) indeed solves
the Maurer--Cartan equation (\ref{MC}).
 Consider first the $d \Omega^{\alpha \beta}$ term in the Maurer-Cartan equation
\begin{eqnarray} \nonumber
d (d X^{\alpha' \beta'} G_{\alpha'}{}^\alpha G_{\beta'}{}^\beta)&=& d X^{\alpha' \beta'} d X^{\gamma \delta}\left (
\left (\frac{\partial  G_{\alpha'}{}^\alpha }{\partial X^{\gamma \delta}} \right)  G_{\beta'}{}^\beta  +
G_{\alpha'}{}^\alpha  \left( \frac{\partial  G_{\beta'}{}^\beta }{\partial X^{\gamma \delta}} \right)
\right) =  \\ \nonumber
&&-  d X^{\alpha' \beta'} d X^{\gamma \delta}G_{\alpha^\prime}{}^\sigma \frac{ \partial G^{-1 \rho}_\sigma}{\partial
X^{\gamma \delta}} G_{\rho}{}^\alpha G_{\beta^\prime}{}^\beta - \\ \nonumber
&&- d X^{\alpha' \beta'} d X^{\gamma \delta}G_{\alpha^\prime}{}^\alpha
G_{\beta^\prime}{}^\sigma \frac{ \partial G^{-1 \rho}_\sigma}{\partial X^{\gamma \delta}} G_{\rho}{}^\beta \,.
\end{eqnarray}
Using
\begin{equation}\label{derivative-1}
\frac{ d X^{\alpha \beta}}{d X^{\gamma \delta}} = \frac{1}{2}(\delta^\alpha_\gamma \delta^\beta_\delta +
\delta^\beta_\gamma \delta^\alpha_\delta) \,,
\end{equation}
and the explicit form of the inverse matrix (\ref{sol2}), one  obtains
\begin{equation}\label{AAA}
\frac{\xi}{4}( d X^{\beta^{\prime} \sigma} G_{\sigma \rho} d X^{\rho \alpha^{\prime}} +
d X^{\alpha^{\prime} \sigma} G_{\sigma \rho} d X^{\rho \beta^{\prime}})G_{\alpha^\prime}{}^\alpha G_{\beta^\prime}{}^\beta \,.
\end{equation}
Let us note, that the product of an even number of $X^{\alpha \beta}$ matrices is antisymmetric  in spinorial indexes, whereas the product of an odd number of $X^{\alpha \beta}$ is a symmetric matrix.
For example,
\begin{equation}\label{PX}
X^{\alpha \gamma} X_{\gamma}{}^{\beta}=- X^{\beta \gamma}X_{\gamma}{}^\alpha,
\quad
X^{\alpha}{}_\gamma X^\gamma{}_\delta X^{\delta \beta}=
+X^\beta{}_\delta X^\delta{}_\gamma X^{\gamma \alpha}, \quad \textrm{etc}.
\end{equation}
As a result,  equation (\ref{AAA}) reduces to
\begin{equation}
\frac{\xi}{2} d X^{\beta^{\prime} \sigma} G_{\sigma \rho}\Bigr|_{\textrm{even}} d X^{\rho \alpha^{\prime}}
G_{\alpha^\prime}{}^\alpha G_{\beta^\prime}{}^\beta=\frac{\xi}{2} d X^{\beta^{\prime} \sigma} G_{[\sigma \rho]} d X^{\rho \alpha^{\prime}}
G_{\alpha^\prime}{}^\alpha G_{\beta^\prime}{}^\beta.
\end{equation}
The evaluation of the second term in the Maurer--Cartan equation  (\ref{MC})  is straightforward with the use of \eqref{103}. After doing so, it is easy to see that  the Cartan form (\ref{sol1})--(\ref{sol3}) solves
the Maurer--Cartan equation (\ref{MC}).


\if{}
\bibliographystyle{utphys}
\bibliography{references}
\end{document}
\fi

\providecommand{\href}[2]{#2}\begingroup\raggedright\endgroup

\end{document}